# Towards the Holodeck: Fully Immersive Virtual Reality Visualisation of Scientific and Engineering Data

Marks, Stefan          Estevez, Javier E.          Connor, Andy M.
Auckland University of Technology
Private Bag 92006, Wellesley Street
Auckland 1142, New Zealand
+64 (9) 921 9999
smarks/jestevez/aconnor@aut.ac.nz

## ABSTRACT
In this paper, we describe the development and operating principles of an immersive virtual reality (VR) visualisation environment that is designed around the use of consumer VR headsets in an existing wide area motion capture suite. We present two case studies in the application areas of visualisation of scientific and engineering data. Each of these case studies utilise a different render engine, namely a custom engine for one case and a commercial game engine for the other. The advantages and appropriateness of each approach are discussed along with suggestions for future work.

## Categories and Subject Descriptors
H.5.1 [**Multimedia Information Systems**]: Artificial, augmented, and virtual realities

## 1. INTRODUCTION
As technology becomes more ubiquitous and immersive, new forms of 'realities' were able to emerge [1]. Exploring concepts such as virtual reality, mixed reality, augmented reality, augmented virtuality, and diminished reality have maintained too high a barrier to entry for any but the most generously funded researchers. All of these new realities merge with or replace parts of the physical world and share common characteristics or goals. As far back as 1913, Edmund Husserl discussed how the artificial world interacts with the physical world of everyday human activities in order to enrich the experiences of perception, affordance and engagement [2]. With modern computing power and the introduction of low cost systems such as the Oculus Rift it is now possible to embody these principles of enrichment in relatively low cost, yet high fidelity systems. This paper outlines the development of a Virtual Reality (VR) facility at Auckland University of Technology and evaluates user experience in two domains, namely engineering and scientific data visualisation.

## 2. BACKGROUND
Virtual reality systems were first explored in the 1960s with the non-interactive Sensorama, Ivan Sutherland's work on interactive computing and head-mounted displays (HMD) around 1965, but became of first real interest in the late 1980s which saw a rapid growth in the development of VR technologies and applications [3, 4]. However, early VR systems often failed to live up to the hype, and were considered to be low-quality and cartoonish. In particular, the visual elements were often jerky and did not respond quickly to the users movements [5]. In addition, very few systems allowed for much active participation in the environment or provided much tactile feedback and as a result the degree of immersion or feeling of presence was low. Some early systems went some way to address these concerns, however surround screen approaches that produced highly immersive environments [6] were exceptionally costly and certainly beyond the reach of all but the most dedicated of research teams, let alone consumers. Whilst these systems did much to bring a range of unique technologies to the attention of a much wider global audience than ever before, they were also responsible for creating a culture of myth, hype and false promise [7]. It is arguable that the VR hype was driven by the booming technology markets of the time, mostly by "internet mania" [8], however whatever the cause, VR was a "roller coaster ride of achievement and failure throughout the 1990s" [7]. Stone goes on to argue that a number of factors contributed to the decline of interest in VR towards the end of the 1990s, which included consistent failures to deliver meaningful and usable intellectual property, expensive and unreliable hardware and an absence of case studies with cost-benefit analyses [7], however the state of the economy post the bursting of the "dot.com bubble" no doubt played a significant role.

Stone [7] continues by arguing that the "serious games" community steadily built up significant momentum since the late 1990s by exploiting the powerful software underpinning video games and that since 2005, serious games have received generally positive outcomes, potentially indicating a rekindling of interest in VR. This is borne out of the large number of publications relating to serious games in the last ten years, with examples in the fields of stroke rehabilitation [9], surgical training simulators [10], supporting cultural heritage [11] and knowledge dissemination [12] to name but a few.

The serious games movement has rekindled interest in VR technologies, but the interest in VR is also growing in many other areas of application. Recent studies have shown the value of immersive VR in the visualisation of scientific, volumetric data [13] and the visualisation of engineering assemblies [14]. However, for VR to not fall foul of its previous promises it is important to address the issues of cost, stability and value that led to its decline in the 1990s.

It has been argued that the combination of consumer-level VR hardware combined with game engines such as Unity 3D[1] can empower hobbyists, professionals, and academics to quickly create VR applications [15]. However the question arises as to whether such approaches can truly deliver the stability and performance required for visualisation of scientific and engineering data.

---
[1] http://www.unity3d.com

To answer this question, this paper will describe the design of an immersive VR visualisation system (Section 3) followed by two preliminary visualisation case studies conducted with the system (Section 4). We will discuss our findings in Section 5, including a discussion of future work.

## 3. THE IMMERSIVE VR SPACE

The implementation of the Immersive VR Space (dubbed the "Holodeck") revolves around several major components, namely a wide area motion capture suite, consumer VR headsets, and several types of render engines (see Figure 1). The motion capture system uses an infrared optical marker based solution. It is currently installed in a room that allows for a capture volume of 6m x 6m x 3m, but can be installed to cover an even larger volume. A total of 24 cameras enable a positional accuracy of up to 0.1mm at 180fps. The cameras are connected via Gigabit Ethernet to a computer with software that captures and processes the location of markers in the space. This information stream is then broadcast into the network by a motion capture server and can be received via wired or wireless connections. For maximum speed and minimum latency, we employ a Gigabit wired network and an 802.11ac wireless access point.

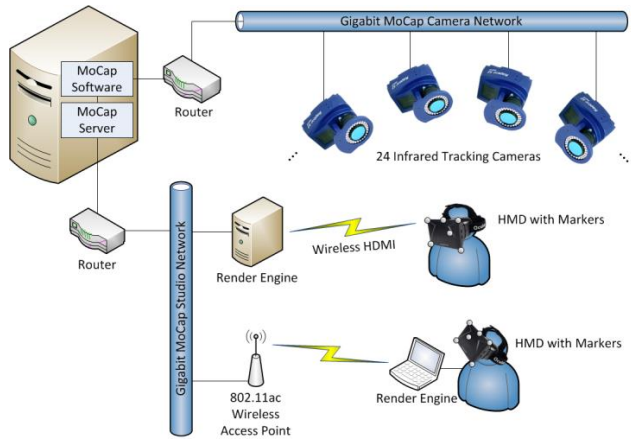

**Figure 1. Block diagram of the Immersive VR Space.**

Computers with render engines use the information provided in this data stream to render scenes based on the positions and orientations of the headsets and other objects like handheld pointers. Several possible combinations of connections and locations of the render engines are possible. One configuration is a mobile computer, possibly carried by the user, wirelessly receiving the position and orientation of the HMD that it renders to directly. A second possible configuration could consist of a fixed computer, connected to the wired network, and wirelessly transmitting a HDMI video signal to a HMD.

One of the major issues in fully immersive VR using HMDs is that of latency between physical motion of the head and the actual change in the visible image according to that motion. "Cue conflict", e.g. a difference in motion perceived by the vestibular organ and as seen by the eyes can easily lead to nausea and other negative physiological effects [16]. The Best Practices Guide of Oculus provides a very useful and exhaustive list of factors that need to be addressed during the design of VR scenarios [17]. One of the most crucial recommendations is to keep the latency below 20ms. We are currently in the process of developing an instrument for measuring the total latency within our system and for optimising the flow of data to reduce the latency.

The VR headset in use is an Oculus Rift Development Kit, version 1 (DK1). Whilst this headset can in itself detect rotational head movement using an accelerometer, it is not capable of detecting linear motion or track the absolute position. The second generation of the DK is using a marker-based tracking system with infrared LEDs on the headset and a single infrared-sensitive webcam to track rotational and translational movement. However, this method is limited to the view frustum of the single camera. To that end, a 3D printed marker mount has been developed that attaches to the headset to enable full positional and rotational tracking. We are using six markers which is more than sufficient to prevent loss of tracking due to occlusion of markers (see Figure 2). By being able to configure the location of some markers, it is possible to have several independently tracked headsets in the motion capture space at the same time, allowing for collaboration between several users.

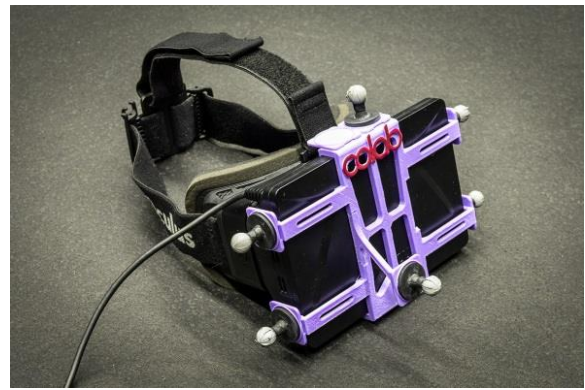

**Figure 2. Headset with markers attached on a custom 3D printed mount.**

To date, a number of case studies have been undertaken to evaluate the performance of the system using different rendering engines and to identify any limitations of the system. These are outlined in the following section.

## 4. CASE STUDIES

### 4.1 Scientific Data Visualisation

The first case study involves the design, implementation and evaluation of a visualisation tool for NeuCube [18], a 3-dimensional spiking neural network, developed by the Knowledge Engineering and Discovery Research Institute (KEDRI)[2].

The number of neurons and connections within NeuCube as well as the 3-dimensional structure requires a visualisation that goes beyond a simple 2D connectivity/weight matrix or an orthographic 45-degree view of the volume. We created a specialised renderer for NeuCube datasets using JOGL (Java Bindings for OpenGL)[3] and GLSL shaders to be able to render up to 1.5 million neurons and their connections with a steady framerate of 60 fps (see Figure 3). In this view, neurons are stylised spheres, and connections are rendered as lines with green colour for excitatory connections and red for inhibitory connections[4].

---

[2] http://www.kedri.aut.ac.nz

[3] http://jogamp.org/jogl/www

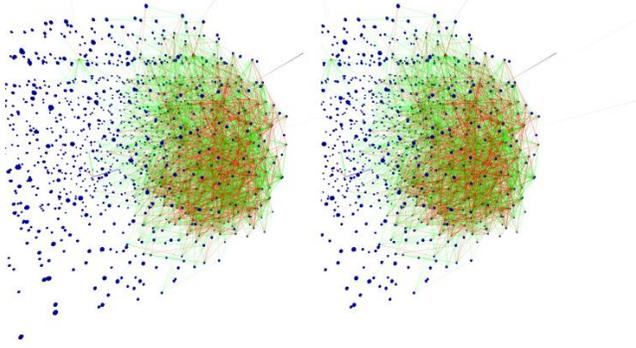

**Figure 3. Stereographic view of NeuCube as being displayed within the HMD of the user.**

In conjunction with the MoCap server, it is possible for the scientists of KEDRI to literally walk through NeuCube and point out individual neurons using a hand-based cursor which will show visual and textual information about the selected neuron, e.g., type, potential, incoming and outgoing connections (see Figures 4 and 5). The network can be simulated in time, showing spiking neuron activity, allowing for slowing down or even pausing the simulation.

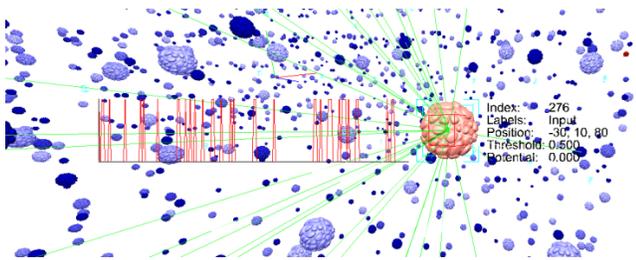

**Figure 4. Cursor node that can be controlled via hand gestures and provides additional information about a specific neuron and its activity.**

In comparison to other scientific visualisation tools for neural networks such as BrainGazer [19] and Neuron Navigator (NNG) [20], our solutions differs in that the user can naturally navigate through the 3D space by simply walking and gesturing instead of using mouse and keyboard shortcuts. Our visualisation is currently very specifically tailored for NeuCube and limited to interactions such as easy information retrieval using the 3D cursor and the animation and playback control for the spiking activity. However, the software can be extended to incorporate larger and more complex datasets such as the brain of the fruitfly drosophila that is being visualised by the two programs mentioned above.

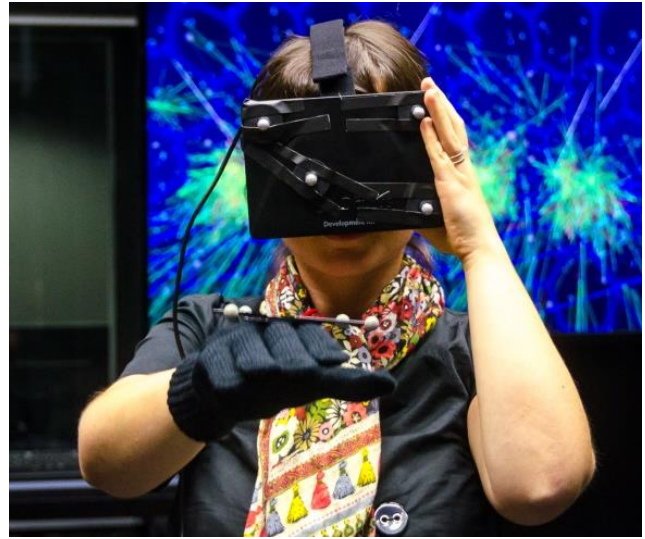

**Figure 5. A user navigating through the virtual representation of the neural network and using an intuitive, hand position based 3D cursor for retrieving information about specific neurons.**

Closer to our application is the work of von Kapri et al, who are using a Computer Assisted Virtual Environment (CAVE) to visualise the spatial structure and activity of a spiking neural network [21]. However, due to the limited space within a cave environment, navigation by simply walking is not possible and requires indirect ways, e.g., by using a controller.

We have not yet conducted a systematic user study, but so far, around 50 visitors of the Immersive VR space have experienced this visualisation. We have observed that, in general, people quickly start to move around and look at structures and point out individual neurons using the 3D cursor. We have received positive feedback about the appearance of the visualisation and that interaction metaphors are very intuitive.

The only larger issue is related to distortion of the field of view and different ratios between actual motion and "visible motion". When users move the hand cursor, experience and muscle memory provides a different "perceived position" of the cursor than is visible in the HMD. This issue, however, can be adjusted by a more careful selection of render parameters and distortion correction for the lenses of the Oculus Rift HMD.

## 4.2 Engineering Data Visualisation

The second case study involved the evaluation of the system in terms of potential for a virtual engineering suite. In particular, there is a focus on the visualisation of CAD data in an immersive manner. Some research in this area has highlighted some of the challenges with applications of this type [22] and one of the goals of this case study was to identify a potential workflow that minimised the noted diversity between CAD and VR models, which have different purposes. In particular we wished to identify whether game engines have the potential to be utilised as a render engine for the final representation derived from the CAD model. Game engines have already gained some popularity in terms of visualisation scientific data [23] and more general VR systems [24], so potential clearly exists for them to be used in serious applications.

---

[4] The white background in Figure 3 is used merely for print purposes. Normally, the background is dark and has a texture to allow for an easier spatial orientation of the user.

Our case study was conducted in conjunction with Stimson Yachts, a New Zealand based yacht design company. Stimson Yachts provided us with a complete CAD model of an 80ft sloop that incorporated the hull form, all interior compartments as well as complete deck rigging. This model was provided in a raw CAD format, in this case a set of Rhinoceros[5] files with different parts of the sloop, e.g., interior, deck, hull, glazing and hardware. The CAD model was first exported with the standard exporter settings to OBJ format, a geometry definition file format first developed by Wavefront Technologies which is open and has been adopted by other 3D graphics application vendors as a universally accepted format. An OBJ import plugin was then utilised to import the model into the 3D editor Blender[6] to pre-process the model for Unity. It became immediately apparent that the large number of polygons in the model (more than 1.5 million triangles) would adversely impact the ability for Unity to maintain a reasonable frame rate. This was worsened because the designer had utilised manufacturers' CAD models for the deck rigging and embedded this in their own design, but the sheer complexity of the CAD model was also a major factor. Another negative factor was the lack of materials defined in the CAD model, which is standard practice in yacht design where the designer passes on the design to an outfitting designer to later specify such requirements.

As a first step towards reduction of the model complexity, the deck hardware was omitted from the export, and the Rhinoceros exporter was configured to the lowest export quality settings. This resulted in a model of 900,000 triangles, which is still a large number for a single object in a game engine, but an improvement to the previous situation. In Blender, the normals of several polygons had to be reversed because they would not show up in the final rendering due to backface culling. The model was then exported to Autodesk FBX, a file format that Unity3D uses natively. After a first render of the model in the game engine, several more polygon- and normal-related problems became apparent and had to be fixed in Blender. The final processing pipeline from the CAD model to the game engine is described in Table 1.

**Table 1. The processing pipeline for 3D CAD models to Unity**

| Step | File Format | Process |
|---|---|---|
| 1 | CAD Model ⇩ OBJ File | Export to OBJ<br>• Choose lowest quality settings for minimum model size |
| 2 | OBJ File ⇩ FBX File | Import to 3D Modeller<br>• Fix surface normals<br>• Improve materials, e.g., by assigning textures, adding transparency, or by setting physical material rendering parameters<br>• Further reduce model complexity<br>• Export to .FBX |
| 3 | FBX File ⇩ Unity3D | Import to Unity3D<br>• Check model and repeat steps 2+3 if necessary |

As a final step, the yacht was placed in a realistic context by the addition of a sky box, ambient ocean sounds, and a realistic Fourier-transform-based ocean surface rendering. A resulting view of the model is shown in Figure 6.

A full experimental evaluation of the performance is planned for the future. The goal of this case study was to determine the feasibility of the approach (see Figure 7). This has been undertaken by considering the opinion of the yacht designer who states[7]:

> "The cues received from the Oculus Rift VR headset are quite persuasive. I was wandering around a simplified 3D model of an 80ft sloop of my design. There was no deck hardware, stanchions or anything beyond the basic deck surfaces at this stage of the project. Yet I could stand in the cockpit, look around and truly sense the geometry of the cockpit seating, the view from the helm to the bow, glance up the rig and check the mainsail trim. I had spent months working on the design and believed myself to be familiar with the form, but I knew from my years in the business that for all the 3D modelling and visualisation on screen, even with high quality 3D renderings, you never actually know what the end result will be until she is in build and you step aboard. So when I found myself looking at the virtual deck with stereoscopic vision I gained a new perspective and an additional level of information that could otherwise only be obtained after construction had started."

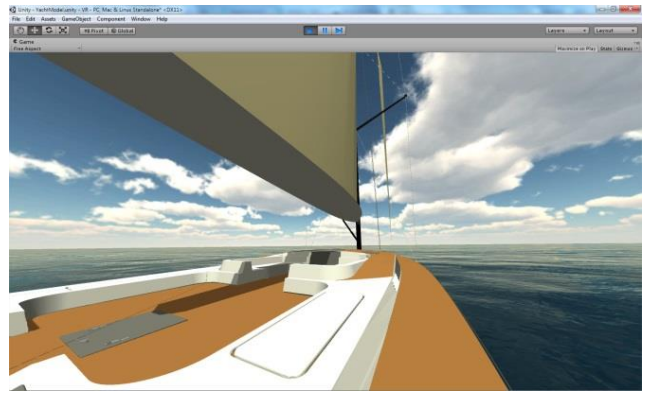

**Figure 6. Screenshot of the yacht model as rendered in the virtual reality.**

Whilst challenges exist in terms of the translation of engineering CAD data into a usable model, the use of a game engine shows promise in terms of providing a useful immersive experience. The possibilities in terms of yacht design include utilising the immersive experience at multiple stages of development. Reducing the number of cameras in the motion capture system would provide the opportunity to develop a low cost, mobile system that can be used in the early design and selling of a yacht.

Such a system would be sufficiently precise to avoid marker occlusion errors yet provide enough precision and fidelity for a potential client to visualise their potential purchase. Later in the design process it is common for a boatyard to build full prototypes based around the intended joinery for the owner to interact with and make design choices. These models are often discarded when the yacht is finally outfitted. Again, a mobile system would allow a more interactive design approach to be deployed: A VR system is used at a boatyard where the full scale model is developed using much lower cost models and the buyer uses the immersive experience to select and change materials that are then rendered in realtime to allow them to quickly evaluate their choices.

---

[5] http://www.rhino3d.com

[6] http://www.blender.org

[7] Designer authored article to appear in Boating NZ Magazine

Such a mobile system offers much greater flexibility than traditional virtual engineering approaches that utilise fixed installations [25], which is particularly useful for the yacht design and boat building industries that are becoming increasingly globalised. The deployment of such a system during the design process will also simplify the workflow in terms of translating the CAD model into the game engine by working with an emerging model rather than a final, complete design.

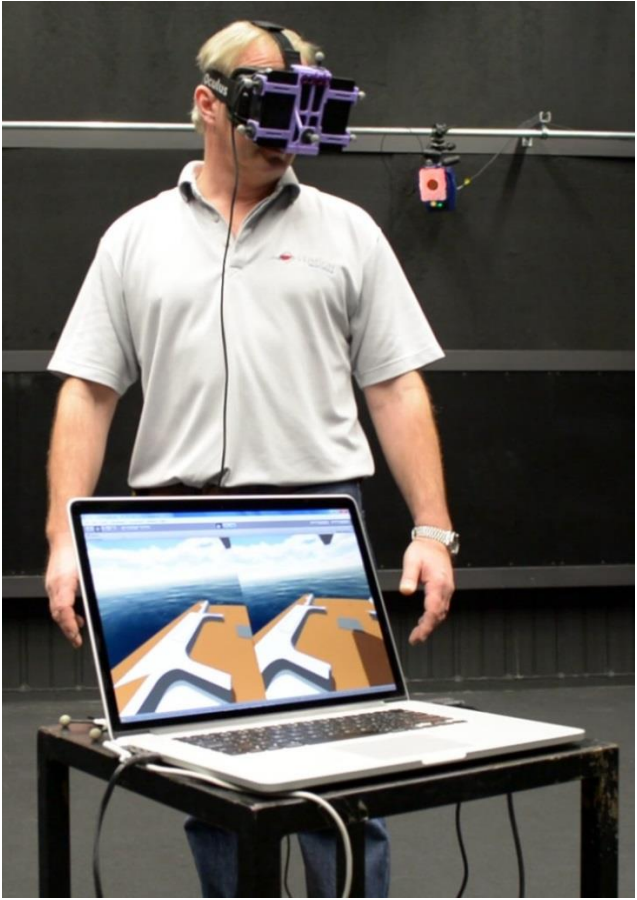

**Figure 7. The yacht designer walking on the virtual representation of his CAD model.**

## 5. DISCUSSION & FUTURE WORK

This paper has briefly outlined the implementation of an immersive suite that combines a consumer VR headset with an existing wide angle motion capture suite. The implementation is designed to be purposefully flexible in order to utilise different network configurations and render engines. Two case studies have been presented that utilise the suite in two different visualisation scenarios. In the first instance, a custom render engine has been written to allow the visualisation of scientific data related to the application of a spiking neural network. The advantages of the custom engine are increased speed of rendering which ensures the level of immersion is sufficient for the purpose. In contrast, the second case study utilizes a sophisticated game engine for the visualisation of engineering CAD data. This has the advantage of greater flexibility but challenges exist in the presentation of CAD data which has significant complexity. This has been addressed through the development of a workflow for simplifying the CAD data in a manner that does not detract from the goal of the visualisation but enables the game engine to provide a useable immersive environment.

The system currently in use is sufficiently mature and robust to demonstrate the potential of a VR based visualisation suite. There are a number of avenues for future work, particularly enhancing the degree of immersion and improving the interactive experience. Our approach already offers significant advantages over similar approaches, namely the ability to navigate the space by walking and to interact with objects through motion tracked objects as opposed to keyboard presses and mouse clicks. However, there is potential to improve the experience further by the introduction a tangible feedback mechanism in both the cases studies considered. As an example, in the scientific data scenario, a low cost haptic glove [26] could replace the virtual hand cursor. It would enable the user to select specific neurons and provide haptic feedback, e.g., to confirm the selection is active or to represent activity of a neuron by vibration.

Further work will also consider the deployment of a mobile system for yacht design, in particular experimentation will be conducted on the minimum number of cameras required to provide an acceptable level of accuracy. This goes hand in hand with experimentation around the maximum number of polygons that can be accommodated to maintain a suitable frame rate which will inform the workflow utilised to import CAD data which will inform the development of an automated workflow to facilitate this process.

An additional consideration for further work will also be systematic analysis of latency in the system as a means to optimize the flow of data and dynamically manage the degree of fidelity in the experience. This will be combined with a formal evaluation of the degree of immersion afforded by the system.

## 6. CONCLUSIONS

This paper has outlined the development and initial evaluation of a virtual reality system for visualising scientific and engineering data. The system utilises an existing motion capture suite that has been outfitted with 24 infrared cameras, however potential exists to reduce this number to further lower costs though this would be accompanied by a reduction in fidelity. Two case studies have been described that investigate the potential for visualisation of scientific and engineering data in an immersive environment. The utilisation of both, specialised render engines and game engines have been considered and both approaches can be deployed, though the complexity of the original data can lead to challenges when importing the data into a useable form.